\documentclass[showpacs,preprintnumbers]{revtex4-1}
\usepackage{graphicx}
\usepackage{bm}
\usepackage{amsmath}

\input{tcilatex}

\begin{document}

\title{Implementation of one-dimensional quantum walks on spin-orbital
angular momentum space of photons}
\author{Pei Zhang,$^{1} $\footnote[1]{zhang.pei@stu.xjtu.edu.cn} Bi-Heng Liu,$^{2}$
 Rui-Feng Liu,$^{1}$ Hong-Rong Li,$^{1}$ Fu-Li Li,$^{1}$ and Guang-Can Guo$^{2}$}

\address{$^{1}$MOE Key Laboratory for Nonequilibrium Synthesis and
Modulation of Condensed Matter, Department of Applied Physics, Xi'an
Jiaotong University, Xi'an 710049, People's Republic of China\\
$^{2}$Key Laboratory of Quantum Information, University of Science and
Technology of China, CAS, Hefei 230026, People's Republic of China}

\begin{abstract}
Photons can carry spin angular momentum (SAM) and orbital angular momentum
(OAM), which can be used to realize a qubit system and a high-dimension
system respectively. This spin-orbital system is very suitable for
implementing one-dimensional discrete-time quantum random walks. We propose
a simple scheme of quantum walks on the spin-orbital angular momentum space
of photons, where photons walk on the infinity OAM space controlled by their
SAM. By employing the recent invention of an optical device, the so-called
'q-plate', our scheme is more simple and efficient than others because there
is no Mach-Zehnder interferometer in the scheme.
\end{abstract}

\maketitle


\section{INTRODUCTION}

Classical random walks (CRW) play an important role in classical algorithms
and have many applications in many realms of science \cite{Barber70}. While
in the quantum worlds, quantum random walks (QRW) are markedly different due
to the phenomenon of interference \cite{Knight03}, which might lead to
extensive applications in quantum information science. The recent
presentation of several quantum algorithms based on QRW has shown the
exponential speed up over classical algorithms \cite{Childs01} as the most
prominent algorithms suggested by Shor \cite{Shor} and Grover \cite{Grover}.
Most recently, Childs has shown that, in principle, any quantum algorithm
can be recast as a QRW algorithm \cite{Childs09}. Meanwhile, numerous
schemes of QRW have been proposed by using various systems, including ion
traps \cite{Travag02}, microwave or optical cavities \cite{Sanders03},
nuclear magnetic resonance (NMR) system\cite{Du03}, optical lattices \cite%
{Dur02}, linear optical elements \cite{Jeong04,Zou06,Do05,Zhang07},
quantum accelerator modes \cite{Ma06} and superconducting circuit quantum
electrodynamics \cite{Xue08}. All these results are advancing forward
towards the ultimate goals of quantum computation.

CRW can take many different forms. For simplicity, we consider the discrete
time random walk on a line. It can be understood as a "walker" standing at
the starting point ($x=0$), holding a coin, decides which side (right or left) he
should take his one step toward by flipping the coin (suppose with head to
right and tail to left). He flips the coin, walks one step to right or left
based on the flipping results. After repeating this process $n$ times, he will
randomly walk to a position $x$. This is a simple model of CRW on a line. We
can get the probability $P_{n}(x)$ of the "walker" being found at $x$ after $%
n$ steps:
\begin{equation}
P_{n}(x) =\frac{1}{2^{n}}\binom{n}{\frac{x+n}{2}}.
\end{equation}
From Eq. 1, we can easily calculate that the mean value is zero and the
standard deviation of this distribution is $\sigma _{c}=\sqrt{n}$, which we
will find being distinctly different to QRW later. QRW occurs on quantum
particles, where their motion described by wave-function and the "coin" changed into "quantum coin" \cite{Kempe03}. Under this version, "quantum walker" does not walk one
step determinately to one position. So there must be interference between
wave-function at different positions. Suppose we define the two level states
as $\left\vert \downarrow \right\rangle $ and $\left\vert \uparrow
\right\rangle $, which correspond to the two sides of "quantum coin", and
one dimensional discrete position $\left\vert x\right\rangle $ ($x$ is
integer) as the walk space. Then we can perform a Hadamard translation ($%
\hat{H}$) to the "quantum coin" as a "quantum coin" tossing. The direction
of "quantum walker" walks to every one step is decided by the results of
tossing "quantum coin", for $\left\vert \downarrow \right\rangle $ steps to $%
\left\vert x-1 \right\rangle $ and $\left\vert \uparrow \right\rangle $
steps to $\left\vert x+1 \right\rangle $. The operator of QRW can be written
as:
\begin{equation}
\hat{U}=e^{i\hat{p}\hat{\sigma} _{z}}\hat{H}.
\end{equation}
Where $\hat{p}$ is the pseudo-momentum operator of the particle confined to
one dimension, and $\hat{\sigma} _{z}$ is the pauli-$z$ operator acting on
the qubit \cite{Travag02}. Then the state of the system after $n$ steps is
\begin{equation}
\left\vert \Psi _{n}\right\rangle =\left( e^{i\hat{p}\hat{\sigma} _{z}}\hat{H%
}\right) ^{n}\left\vert \Psi _{in}\right\rangle
\end{equation}%
For different initial states of "quantum coin", the results of QRW are very
different. We can also calculate the mean value which is nonzero and the
standard deviation of the probability distributions $\sigma _{q}\approx\frac{%
\sqrt{2}}{2}t$ \cite{Cart03}. These are two main different features between
CRW and QRW.

Photons are well known to be extremely powerful for carrying and
manipulating quantum information. QRW have been experimentally demonstrated
in linear optical systems \cite{Do05, Zhang07}. Ref. \cite{Do05} realized a
QRW by employing photon's polarizations and space modes. But the number of
linear optical elements grows quickly when steps increasing and the results
are not very perfect. Ref. \cite{Zhang07} carried out an experiment of QRW
by using photon's up-down routes and orbital angular momentum (OAM). This
setup is simpler than \cite{Do05} because there are only two space modes and
the photons "walk" on OAM space. However, these two experiments both have a
critical shortcoming, that is, they both need Mach-Zehnder interferometers.
The instability of interferometer directly leads the precision decreasing
when steps of random walk increasing. In this paper, we design an
experimental scheme of QRW by using linear optical system. We employ
photon's spin angular momentum (SAM) as "quantum coin" and OAM space as
one-dimensional walk space. By using q-plates \cite{Mar}, our scheme is very
simple and efficient for there is no Mach-Zehnder interferometer in the setup. This
paper is organized as follow: first we briefly introduce the OAM of photons
and the q-plates; then we will give a detailed description and discussion of
our experimental scheme; last is the conclusion.

\section{ORBITAL ANGULAR MOMENTUM and Q-PLATE}

As we know, light beams have polarizations and helical modes. For photons,
it is corresponding to spin angular momentum (SAM) and orbital angular
momentum (OAM), respectively. SAM is associated with the circular
polarization of light, each photon carries $\pm\hbar$ angular momentum
depending on the handedness of the polarization. OAM is associated with
transverse amplitude and phase profile in the cross section orthogonal to
the propagation axis, and one photon carries an intrinsic orbital angular
momentum given by $m\hbar $ if the wave function contains phase term $%
exp(im\phi )$ \cite{Allen92}. Since the quantum nature of OAM was
demonstrated by Mair et al. \cite{Mair01}, more and more studies have
focused on encoding the high-dimensional quantum information by using OAM of
a single photon \cite{OAM}. These works open up a new possibility for
generation and manipulation of higher-dimensional quantum states and
implementation of higher-dimensional quantum information processing
protocols, by using OAM as a quantum information carrier.

When light propagates through a medium, the SAM and OAM are not always
simultaneous independent coupling with the matter \cite{He}. In vacuum or in a
homogeneous and isotropic medium, the SAM and OAM are separately conserved.
But in a medium which is inhomogeneous and anisotropic, they are not. In
appropriate condition, the exchange with matter of the SAM affects the
direction of the exchange of the OAM, and it is possible that the angular
momentum only converting between SAM and OAM. A device called q-plate is used to
conduct this function. Q-plate is a slab of liquid crystal (or uniaxial
birefringent medium) with special structure, and it is essentially a
retardation wave plate whose optical axis is aligned nonhomogeneously in the
transverse plane in order to create a topological charge $q$ in its
orientation. If we assume $q=1$, for example, each photon being converted
from right circular to left circular change its spin z-component angular
momentum from $-\hbar $ to $+\hbar $, and the orbital z-component angular
momentum changes $-2\hbar $. Therefore, the total variation of angular
momentum is nil, and there is no net transfer of angular momentum
to the q plate. The plate acts only as a medium for the conversion between
spin and orbital angular momentum. If $q\neq 1$, it will exchange an angular
momentum of $\pm 2\hbar (q-1)$ with each photon. Using Dirac marks, suppose
the initial state is $|\uparrow _{S},m_{L}\rangle $ or $|\downarrow
_{S},m_{L}\rangle $, which refers to the photon carrying SAM with $\hbar $
(left circular polarization) or $-\hbar $ (right circular polarization) and
OAM with $m\hbar $. The action of a tuned q-plate of charge $q=1$ on this
state can be summarized as follows:
\begin{equation}
(|\uparrow _{S},m_{L}\rangle ,|\downarrow _{S},m_{L}\rangle )\overset{q-plate%
}{\longrightarrow }(|\downarrow _{S}(m+2)_{L}\rangle ,|\uparrow
_{S},(m-2)_{L}\rangle )
\end{equation}%
In this way, a photon in an OAM state $m=0$ is transformed into $m=\pm 2$,
or a photon in $m=\pm 2$ to $m=\pm 4$ or $m=0$, depending on the
polarization. This spin-controlled-orbital transform is very suitable for
implementing quantum walks. Here we use the character of infinity space of
OAM and the properties of q-plates to plan out an implementation of
one-dimensional discrete time QRW, where SAM refer to the quantum "coin" and
$m$ of OAM refer to the photon "positions" when random walk.
We will give a detailed explanation later.

\begin{figure}[tbh]
\includegraphics[width=10cm]{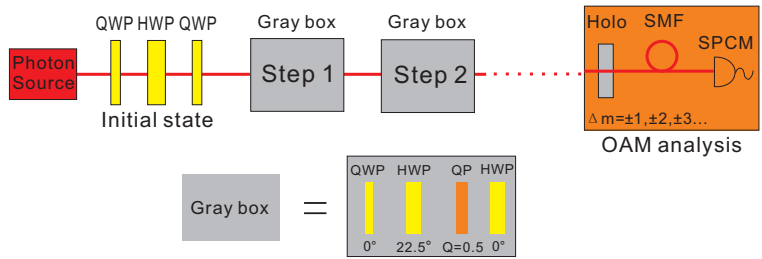}
\caption{Experimental scheme of one-dimensional quantum walks on
spin-orbital angular momentum space of photons. One half wave plate (HWP)
sandwiched by two quarter wave plates (HWP) can realize any single bit
unitary operation, which is used as a initial state preparation. Gray box,
which is made up by two HWPs, one QWP and one q-plate (QP), performs the QRW
operation $\hat{U}$. An n-steps QRW can be achieved by applying $n$ times of
gray box. At end of the setup is the OAM analysis system, which contains a
computer-generated hologram (Holo), a single mode fiber (SMF) and a single
photon counting module (SPCM). When we change different holograms (with
different $\Delta m$), we can get the probabilities of the photons being in
each OAM state. }
\end{figure}

\section{EXPERIMENTAL SCHEME}

QRW on the OAM space of photons have been studied both theoretically \cite%
{Zou06} and experimentally \cite{Zhang07}. However, the steps of QRW can
hardly be increased in the experiment, because of the stability of
Mach-Zehnder interferometer decreasing quickly when the arms of
interferometer increasing. We plan out a QRW scheme without interferometers
which can achieve a multi-step QRW stably and can be easily implemented in
the lab. In our scheme, we can use the q-plates with $q=0.5$ \cite{Exp1}
where the OAM index $m$ changes to $m\pm1$ when passing it.

Our experimental scheme is shown in Fig. 1. The single photon source (can be
realized by deep attenuating the coherent laser or by nonlinear process)
emits one photon with linear polarization (suppose horizontal polarization)
and OAM $m=0$. The state can be writen as $|H,0\rangle =|(R+L)/\sqrt{2}%
,0\rangle $, where H is short for horizontal polarization, R and L are short
for right circle polarization and left circle polarization. We can also use
a set of wave plates made by one half wave plate (HWP) sandwiched by two
quarter wave plates (QWP) \cite{Englert}, to prepare different initial polarization
states as quantum coin states if needed. Then we will toss the quantum coin
and decide which direction should step to. This process is achieved by the
gray box shown in Fig. 1. Hadamard gate ($\hat{H}$) is a kind of quantum
coin tossing, which can realize translation as following:
\begin{eqnarray}
&&\text{ \ \ \ }\hat{H}|R\rangle =\frac{1}{\sqrt{2}}(|R\rangle +|L\rangle )
\\
&&\text{\ \ \ \ }\hat{H}|L\rangle =\frac{1}{\sqrt{2}}(|R\rangle -|L\rangle ).
\notag
\end{eqnarray}%
In the gray box, a QWP at $0^\circ$ and a HWP at $22.5^\circ$ make up a $%
\hat{H}$ gate for $|R\rangle $ and $|L\rangle $ states. After flipping the
coin, then the photon will step to $m+1$ when coin state is $|R\rangle $ or
to $m-1$ when coin state is $|L\rangle $. Here the photon's QRW positions
are denoted by the OAM numbers of photon. From Eq. (4), we know a q-plate
can easily realize this process combining with a HWP at $0^\circ$ for spin
reverse. So, the QRW operator $\hat{U}$ is achieved by the gray box. Then we
can apply this operation $n$ times as a $n$ steps QRW. If the initial state
is $\left\vert \Psi _{in}\right\rangle =|R,0\rangle $, we can deduce three
steps for example:
\begin{eqnarray}
&&\left\vert \Psi _{in}\right\rangle \overset{\hat{U}}{\longrightarrow }%
\frac{1}{\sqrt{2}}\left( \left\vert R,-1\right\rangle +\left\vert
L,1\right\rangle \right) \\
&&\text{ \ \ \ }\overset{\hat{U}}{\longrightarrow }\frac{1}{2}\left(
\left\vert R,-2\right\rangle +\left\vert R,0\right\rangle +\left\vert
L,0\right\rangle -\left\vert L,2\right\rangle \right)  \notag \\
&&\text{\ \ \ \ }\overset{\hat{U}}{\longrightarrow }\frac{1}{2\sqrt{2}}%
\left( \left\vert R,-3\right\rangle +\left\vert L,-1\right\rangle +2\left\vert
R,-1\right\rangle -\left\vert R,1\right\rangle +\left\vert L,3\right\rangle
\right).  \notag
\end{eqnarray}%
From the third step, we can see the distribution of probabilities of photons
being found at different positions (photon's OAM) is asymmetric, which is
very different to CRW. At the end of this setup, we should detect the
probabilities of the photons being in a OAM state $|m\rangle $. This is
achieved by using different computer-generated holograms (Holo) \cite{Bazh90} with $\Delta
m=\pm 1,\pm 2,\cdots,\pm n$ and a single mode fiber (SMF). We know that the
single mode fiber can only pass $TEM_{00}$ mode light and has a strong
absorption to other higher order modes. So, to detect photon in OAM state $%
|m\rangle $, we can use a hologram of $\Delta m=-m$ to change the photon
into OAM state $|0\rangle $, which then can pass through the SMF and
detected by a single photon count modulator (SPCM).

The efficiency of our setup is
very high due to high transmitted efficiency of q-plate \cite{Mar} and linear
optical elements. We know that the
different OAM photons will spatially separate in their radial pattern by
free propagation. This may induce imperfection when take a long free
propagation. However, this can be resolved by placing the q-plates in the
common focal plane of two lens (the lens are not appeared in our setup) \cite{Slu09}. It is might be inconvenient in changing the different
holograms for position detecting. For $n$ steps QRW, the detection needs to
change $2n$ different holograms. But this is not very hard to our scheme,
because it can keep a long time stable for all the different OAM modes are
in one spatial mode and no use of Mach-Zehnder interferometer.

There is another work of experimental realization of QRW by taking advantage of OAM of photons as walk space (see Ref. \cite{Zhang07}). Comparing with the scheme of Ref. \cite{Zhang07}, our scheme has two main improvements: the
first one is the efficiency, because the transmitted efficiency of q-plate is higher than computer-generated holograms;
the second is the stability, our scheme is inherently stable for no need of Mach-Zehnder interferometer, because we know that Mach-Zehnder interferometer is very sensitive with environment especially
for long arms. So, by employing SAM as quantum coin and using
q-plate, as we analyzed above, the scheme of this paper is more efficient and
simple. QRW on a line can be understood as an interference phenomenon
\cite{Knight03}, although there is no Mach-Zehnder interferometer in our
scheme, there still exists interference. From Eq. (6), we can see that in
the third step of QRW, the spin interference between state $\left\vert
R,0\right\rangle $ and $\left\vert L,0\right\rangle $ occurs on the third
Hadamard gate.

\section{CONCLUSION}

By employing photons SAM and OAM, we put forward a simple and efficient
scheme of discrete-time QRW on a line. We have introduced an optical device,
the gray box, by which the QRW operation $\hat{U}$ can be realized. By applying
$n$ times of gray box, it is very simple to realize a $n$ steps QRW. The efficiency of
our scheme is high and the stability is very good, which are suitable for
implementing multi-step QRW. We think our scheme is very simple to do some
QRW algorithm experimentally.

\section{ACKNOWLEDGEMENT}

The authors thank Professor L. Marrucci for helpful
discussion on q-plates, and acknowledge the National Fundamental Research Program (2010CB923102),
National Natural Science Foundation of China (Grant
No. 60778021, 10674106 and 10774117) for supporting this work.


\begin{thebibliography}{99}
\bibitem{Exp1} Generally, q-plate with $q=1$ is easy to manufacture and
widely used in the former articles. Here we can also use $q=1$, which can
realize a QRW on a $even$ $m$ OAM space.
\end{thebibliography}
\end{document}